\documentclass[12pt]{article}

\usepackage[utf8]{inputenc}
\usepackage{multirow, amsmath, biblatex, titling, breqn}
    \usepackage[colorlinks]{hyperref}
    \hypersetup{
    colorlinks=true,
    linkcolor=blue,
    filecolor=magenta,      
    urlcolor=blue,
}

\usepackage[inline]{enumitem}
\usepackage{tgtermes, hyperref} 
\usepackage[table]{xcolor}  
\usepackage{multicol, bm,tikz,tabularx,ragged2e,booktabs,caption,float, lscape, lipsum, setspace}

\usepackage[T1]{fontenc}
\usepackage[margin = 1in]{geometry}

\begin{document}

\title{Incorporating baseline covariates to validate surrogate endpoints with a constant biomarker under control arm}
\date{}
\author{Emily Roberts$^{1*}$, Michael Elliott$^{1,2}$, Jeremy M. G. Taylor$^1$\\
$^1$Department of Biostatistics, University Michigan\\
1415 Washington Heights Ann Arbor, MI 48109\\
ekrobe@umich.edu\\
$^2$ Survey Methodology Program, Institute for Social Research Ann Arbor, MI 48109
}

\maketitle
\vspace{-.7in}
\subsection*{Abstract}

A surrogate endpoint $S$ in a clinical trial is an outcome that may be measured earlier or more easily than the true outcome of interest $T$. In this work, we extend causal inference approaches to validate such a surrogate using potential outcomes. The causal association paradigm assesses the relationship of the treatment effect on the surrogate with the treatment effect on the true endpoint. Using the principal surrogacy criteria, we utilize the joint conditional distribution of the potential outcomes $T$, given the potential outcomes $S$.
In particular, our setting of interest allows us to assume the surrogate under the placebo, $S(0)$, is zero-valued, and we incorporate baseline covariates in the setting of normally-distributed endpoints. We develop Bayesian methods to incorporate conditional independence and other modeling assumptions and explore their impact on the assessment of surrogacy. We demonstrate our approach via simulation and data that mimics an ongoing study of a muscular dystrophy gene therapy.

\subsection*{Keywords} Bayesian methods,  principal stratification, subgroup effects, surrogate endpoints 

\newpage

\section{Introduction}

\vspace{0.5cm}

Although randomized clinical trials are largely considered the gold standard to evaluate treatment efficacy, methods that lower trial cost and shorten the length of the study are often sought after in the medical field. Surrogate endpoints $S$ are biologically plausible intermediate outcomes that are strongly related to the true outcome of interest $T$. In a trial, these endpoints may be measured earlier or more effectively to quickly disperse treatments to patients. Popular examples of potential surrogate endpoints include CD4 blood counts for HIV mortality and immune responses for vaccine efficacy. It is crucial to collect data and validate such an endpoint before using in a large-scale trial.

Prentice's landmark 1989 paper proposed criteria in a single trial setting to evaluate statistical surrogates: that the surrogate $S$ both be related to the outcome and that it captures the effect of the treatment $Z$ on $T$.$^1$ Other criteria have since been proposed, as it has been shown simple criteria may not ensure a seemingly useful surrogate will predict a beneficial treatment effect.$^2$ $S(z)$ and $T(z)$ refer to the endpoint values had the treatment, possibly counter-factually, been assigned to level $z$. Since $S$ is measured after treatment assignment, conditioning on the surrogate distorts the causal pathway and interpretation of the treatment effect in a regression model. Using Rubin's potential outcome causal framework,$^{3,4}$ principal surrogacy proposed the solution of using both potential intermediate outcomes by considering the surrogate values under each treatment as pre-treatment variables.$^5$ Since both surrogate outcomes are hypothetically determined prior to randomization, surrogates can be evaluated based on principal causal effects defined on the distribution of $T(1) - T(0)$ conditional on principal strata defined by $S(1) - S(0)$. Gilbert and Hudgens (forward as GH$^6$) elaborated on causal necessity, as proposed by Frangakis and Rubin, by terming average causal necessity and average causal sufficiency. These require that there be no average effect of the treatment on the true outcome in the strata where there is no average effect on the surrogate, and similarly that there exists an average treatment effect on the true outcome in the strata where is there an average effect on the surrogate. Further, they define the Causal Effect Predictiveness (CEP) curve as a visualization of both requirements across the value of the surrogate endpoints $(s_1, s_0)$.

Our work is motivated by an ongoing study of a muscular dystrophy treatment.$^7$ In this trial of a new gene transfer therapy, patients received a micro-dystrophin transgene to produce the micro-dystrophin protein. The potential surrogate $S$ is micro-dystrophin expression as measured by western blot methods, and the primary outcome of interest $T$ is the North Star Ambulatory Assessment (NSAA) functional score on a continuous scale. $S$ is measured at only one time point, while $T$ is measured before randomization as well as after the gene transfer therapy. Since muscular growth and deterioration due to the disease have major impact on physical movement during childhood, both baseline ambulatory ability and age are important to take into consideration. In this setting, patients do not produce significant amounts of micro-dystrophin protein at baseline, so it can be assumed that the value of the surrogate under placebo is approximately equal to 0. This scenario where $S(0)$ is fixed to 0 also commonly arises in vaccine efficacy studies. Since those in the placebo group necessarily have no immune response without the vaccine antigens, GH refer to this simplified setting as the constant biomarker case.

Quantities related to vaccine efficacy were developed in the HIV and pertussis settings,$^{8,9,10}$ and GH were among the first to formalize the surrogate validation methodology in a HIV vaccine efficacy trial. Still, a major challenge of characterizing these causal effect summaries is dealing with nonidentified parameters arising from use of potential outcomes, so Follmann$^{12}$ suggested the closeout placebo vaccination design to avoid the unobserved outcomes. GH focused on modeling assumptions to identify the causal quantities and generalized previous work of the baseline immunogenicity predictor (BIP) $W$ to estimate the missing $S(z)$ value. Related work has proposed augmented trial design ideas such as the baseline surrogate measure and the cross-over design, respectively, and other authors have imposed conditional independence assumptions of the outcomes$^{13}$. Subsequently, several authors have addressed particular models and designs for vaccines and immune correlates of protection.$^{14,15,16,17,18}$

Under the general Bayesian paradigm, with specific assumptions about parameter values,  both Zigler and Belin$^{19}$ and Conlon, Taylor, and Elliott$^{20}$ proposed to consider the full joint distribution of potential outcomes to create the CEP curve using imputation strategies applicable to settings beyond the constant biomarker case. Work in the frequentist setting by Alonso, Van der Elst, and Meyvisch utilized potential outcomes and the information-theoretic framework to propose a surrogate predictive function with a two-step procedure for dealing with non-identifiability.$^{21}$ Making the constant biomarker assumption results in fewer missing potential outcomes to impute, which allows us to focus on the sensitivity of modeling assumptions. In this work, we build upon previous models of the joint distribution of potential outcomes $S(0),\ S(1), \ T(0),\ T(1)$ by exploring trial and modeling considerations when controlling only three of these endpoints as applicable to our motivating clinical trial. $^{20,22}$

We propose techniques and design considerations with the goal of achieving gains in estimation efficiency. Since the surrogate and true outcome values are only observed for the assigned treatment, we consider the counterfactual outcomes as missing data and implement an imputation strategy for estimation. We compare this algorithm to instead using only observed data and prior distributions for nonidentified correlation parameters. The novelty of this work is the incorporation of baseline covariates with two objectives in mind: first, conditioning on baseline covariates may improve the plausibility of conditional independence assumptions, and second, it allows us to make inference about whether there are subgroups of the population for whom the quality of the surrogate varies. For the latter, we propose to stratify the previously marginal estimands for validation by conditioning on patient characteristics. In our application, we focus on one such example of a baseline covariate, namely that of the true outcome measured pre-treatment, which is similar to the BSM proposed by other authors. This particular measurement allows for multiple definitions of the true outcome of interest and other trial design decisions. We recognize that by viewing the baseline as a noisy estimate of $T(0)$ similar to a measurement error problem, it could provide improved identifiability or yield more informative prior distributions for nonidentified parameters. 

In Section 2, we propose the model and incorporation of baseline covariates in the surrogate setting. We define the conditional surrogacy validation metrics and suggest potential reasons to use the covariates such as to make conditional independence assumptions and raise consideration for how to define the trial endpoint. In Section 3, we describe the proposed Bayesian estimation methods using either an imputation scheme or observed data algorithm. Simulation studies are shown in Section 4, and our data example is explored in Section 5 before a concluding discussion in Section 6.
\vspace{-.2in}

\section{The Model}

Using the causal association framework, we first consider the joint distribution of three continuous potential outcomes under a binary treatment $Z$. Since $S(0) = 0$, we assume a multivariate normal distribution of the counterfactual surrogate and true outcomes for each subject:
\begin{equation}
\left(
\begin{array}{c}
S(1) \\ T(0) \\ T(1) 
\end{array}
\right) \sim MVN \left( \left( \begin{array}{c}
\delta_1\\
\delta_2 \\
\delta_3
\end{array}\right) , 
\left( \begin{array}{ccc}
\sigma_{S1}^2 &\rho_{10}\sigma_{S1}\sigma_{T0} &\rho_{11}\sigma_{S1}\sigma_{T1} \\
&\sigma_{T0}^2 &\rho_T\sigma_{T0}\sigma_{T1} \\
&&\sigma_{T1}^2 \\
\end{array}\right) \right) 
\end{equation}

\subsection{Assumptions}

In this setting, we focus on clinical trial scenarios where we can assume $S(0) = 0$ (the methods can be extended to more general settings). The causal inference assumptions we make are the Stable Unit Treatment Values Assumption and ignorable treatment assignment, meaning the potential outcomes for any individual do not vary with the treatments assigned to other individuals, and treatment assignment is independent of potential outcomes conditional on all covariates ($P(Z=1|T(0), T(1),X)  = P(Z=1|X)$), respectively. Since we do not observe combinations of joint outcomes $\{T(0), S(1)\}$ or $\{T(0), T(1)\},$ the correlation parameters $ \rho_{10},\rho_T$ are not identified. We will consider various approaches to obtain identifiability through the use of proper priors, conditional independence assumptions, and/or fixing unknown parameters via sensitivity analyses, and consider how our models may be adapted if additional baseline data is available. Then we use the specified joint model to impute the missing counterfactual values. This potential outcomes approach captures the causal associations for validation. 

The multivariate normality assumption provides many convenient results by being analytically tractable and allowing for closed form quantities for surrogate validation which we describe below. As this may not hold in practice, we later verify the sensitivity of this distributional assumption and assess the robustness of the results in the presence of model misspecification.
Other work has incorporated copula models for non-normal data.$^{23}$ 
\vspace{-.2in}

\subsection{Baseline Covariates}

Some estimates for surrogacy quality have wide confidence bands that make definitive recommendations difficult. While most causal metrics are reported marginally, it may be beneficial to use baseline covariates $X$ in the analysis. One interest is to assess effect modification: that is, if there exist subgroups of patients to determine for whom the surrogate will work particularly well for the true outcome, such as for males or those who are young. This has the potential to reduce the risk of observing the surrogate fail in a certain patient population after approval for use in subsequent trials. It is also possible that covariates would help predict membership of principal strata. $X$ may explain dependence or confounding between $S$ and $T$ that may occur in finite samples, even after trial randomization. Statistical benefits may be seen in the estimation accuracy as well via more accurate imputation of the missing counterfactual values to both reduce bias and gain efficiency. Finally, conditioning on $X$ might be expected to reduce correlations amongst the potential outcomes, allowing us to make stronger conditional independence statements after conditioning on baseline covariates. Note that these latter examples' effects can be reported as either conditional on $X$, or integrated over the empirical distributions of covariates to provide marginal estimates.

The conditional model can be written with effects of $X$ in the mean structure (therefore the parameters in this covariance structure differ and $\theta$ represents the conditional correlations)
\vspace{-.2in}

\begin{equation}
\left(
\begin{array}{c|c}
S(1) & \\ T(0)&X\\ T(1) 
\end{array}
\right) \sim N \left( \left( \begin{array}{c}
\omega_1 +\omega_2 X\\
\omega_3+\omega_4 X\\
\omega_5 +\omega_6 X
\end{array} \right), 
\left( \begin{array}{ccc}
\epsilon_{S1}^2 & \theta_{10} \epsilon_{S1} \epsilon_{T0} &  \theta_{11}\epsilon_{S1} \epsilon_{T1}  \\
&\epsilon_{T0}^2 & \theta_{T}\epsilon_{T0} \epsilon_{T1}  \\
&&\epsilon_{3}^2\\
\end{array}\right)
 \right) 
\end{equation}

We note that this model still has two nonidentified parameters, $\theta_{10}$ and $\theta_T$. In the above model $X$ is a scalar, but it could be generalized to a vector. Furthermore, one of the $X$ components might be known to be highly related to either $S$ or $T$ (such as a pre-treatment measurement). In these specific cases that provide additional information, it may be feasible to make further assumptions about the model structure.
\vspace{-.2in}
\subsection{Surrogacy Validation}

The validation causal quantities derived from conditioning on strata of the surrogate can be written as a function of the model parameters. These quantities can be viewed graphically in the causal effect predictiveness (CEP) surface as a line with intercept and slope based on causal effects as the difference in surrogate potential outcomes $S(1)-S(0)=s$ on the $x$-axis and difference in the expected, conditional true outcomes $E(T(1)-T(0)|S(1)-S(0)=s)$ on the $y$-axis. In the case of Gaussian distributions, as in equation 1, $E(T(1)-T(0)|S(1)-S(0)=s)$ is linear in $s$ and has the form  $=\gamma_0+\gamma_1 s$. By displaying expected change in potential outcomes, conditional on the actual surrogate change $s$, the plots demonstrate if the surrogate is valid, meaning small (large) causal effects on a surrogate are associated with small (large) causal effects on the outcome. When the distribution of outcomes is multivariate normal, average causal necessity and  average causal sufficiency are fulfilled if 
$\gamma_0=0$, the expected change in true outcome when there is no change in the surrogate outcome at the origin, and $\gamma_1 \neq 0$, the expected change in true outcome when there is a nonzero change in the surrogate outcome. Under the multivariate normal distribution in equation 1, these values from the conditional expectation can be written as
\vspace{-.5cm}   $$\gamma_0 = (\mu_{T1}-\mu_{T0})- \gamma_1 \mu_{S1} = (\delta_3 - \delta_2)- \gamma_1 \delta_1 \ \ \ \ \ \ \ \ \ \gamma_1 = \frac{\rho_{11}\sigma_{T1} -\rho_{10}\sigma_{T0}}{\sigma_{S1}}$$\vspace{-1.4cm}

\noindent or when incorporating baseline covariates, $\gamma_0=(\mu_{T1|X}-\mu_{T0|X})-\gamma_1\mu_{S1|X}$ where $\gamma_1 = \frac{\theta_{11}\epsilon_{T1} -\theta_{10}\epsilon_{T0}}{\epsilon_{S1}}$. Our goal is to estimate all parameters in the distribution so we can calculate $\gamma_0$ and $\gamma_1$ and determine if $S$ is a valid surrogate for $T$. We can understand $\gamma_1$ by rewriting the quantity as $\frac{\omega_{11} - \omega_{10}}{\sigma_{S1}}$ and can consider when $\omega_{11} > \omega_{10}$ that the slope will be positive, and when the ratio $\omega_{11}/\omega_{10}$ is larger, the magnitude of $\gamma_1$ is larger. Another way to look at this term is as $\frac{Cov(T(1), S(1)) - Cov(T(0), S(1))}{Var(S(1))}$, where the sign and magnitude of $\gamma_1$ is determined by the covariances between $S(1)$ and the true outcomes.

For scenarios where we believe the surrogate works particularly poorly for certain patient groups, we would be interested in a stratified CEP curve. In other settings, we may simply incorporate covariates as an intermediate step while remaining interested in the marginal CEP curve. To denote the difference, let $\gamma_{0, C}$ and $\gamma_{1, C}$  correspond to a model fit using $X$, compared to the marginalized estimates that is accomplished by empirically averaging over the distribution of $X$. Later, we will further differentiate these respective conditional $C$ and marginal $M$ models based on how we define the outcome. From equation 1, $T(1)-T(0)|S(1),x$ has a normal distribution, so the expected value will be written as $\gamma_{0, C} + \gamma_{1, C} s$. In this expression $\gamma_{0, C}$ can depend on the covariate $X$, but by the assumptions of the model $\gamma_{1, C}$ (i.e. the covariance) does not depend on $X$.  From these values, the marginalization is written
\vspace{-.3cm} \begin{equation}\int_x E(T(1)-T(0)|S(1),X=x)f(X|S(1)=s)dx = \int_x \frac{E(T(1)-T(0)|S(1),x)f(S(1)|x)f(X)}{f(S(1))}dx
\end{equation}
using Bayes rule (see Web Appendix A). Once we obtain these quantities, we plot the marginal effect over values of $s$ and summarize it by fitting a linear model to estimate $\gamma_{0,M}$ and $\gamma_{1, M}$.
\vspace{-.3in}
\subsection{Subgroups in Surrogacy and Treatment Effects}

Based on our definition of $\gamma_0,$ our concept of surrogacy subgroups is analogous to heterogeneous treatment effects existing. Since the quantity $\gamma_{0,C}$ depends on $x$, a surrogacy subgroup effect will occur only when there is an interaction with the treatment effect on either the surrogate or the true outcome. In contrast, we are not assuming the absence of unobserved heterogeneity in the sense of a sharp null existing.$^{24}$

\vspace{-.5cm} \subsection{Conditional Independence}

Conditional independence assumptions are frequently made in causal inference and for surrogate endpoint validation in particular (see Conlon et al., for examples of assumptions made)$^{18,25,26,6}$. Briefly, when $S(0)$ is not restricted to the value of 0, common assumptions among the endpoints include strong statements about the outcomes such as 1) $T(0) \perp T(1) | S(0), S(1)$; 2) $S(0) \perp T(1) | S(1)$; 3) $S(1) \perp T(0) | S(0)$; and intuitively weaker 4) $S(0) \perp T(1) | S(1), T(0)$ and 5) $S(1) \perp T(0) | S(0), T(1)$.

In our setting, the reasonable assumption that is still feasible is a collapsed version of 5: $S(1) \perp T(0) | T(1)$. Determining if a conditional independence assumption is plausible is context-dependent. Examining the assumption, it is not unreasonable to believe that given the true outcome under $Z=1$, the surrogate for $Z=1$ is independent of the true outcome under the opposing treatment. This assumption implies the relationship among equation 2 parameters that $\theta_T = \theta_{10}/\theta_{11}$. In the multivariate normal setting, this combination of correlation parameters on the joint normal scale is derived  by setting a conditional covariance term to $0$ (described in Web Appendix A),  which corresponds to a zero element in the inverse of the correlation matrix. 
After adjusting for baseline covariates $X$, conditional independence of the outcomes may be more likely to hold, which in turn reduces the number of parameters to estimate by one (a restatement of equation 2 in which $\theta_{10}$ has been replaced by $\theta_{11} \times \theta_{T}$):

\begin{equation}
\left(
\begin{array}{c|c}
S(1) &\\ T(0) & X\\ T(1) 
\end{array}
\right) \sim N \left( \left( \begin{array}{c}
\omega_1 +\omega_2 X\\
\omega_3+\omega_4 X\\
\omega_5 +\omega_6 X
\end{array} \right), 
\left( \begin{array}{ccc}
\epsilon_{S1}^2 &\theta_{11}\theta_T\epsilon_{S1}\epsilon_{T0} &\theta_{11}\epsilon_{S1}\epsilon_{T1} \\
&\epsilon_{T0}^2 &\theta_T\epsilon_{T0}\epsilon_{T1} \\
&&\epsilon_{T1}^2 \\
\end{array}\right)
 \right) 
 \end{equation}

\noindent This potentially increases efficiency and helps with identifiability since there is only one nonidentified parameter ($\theta_T$).

\subsection{Design Considerations and Defining the Endpoint}

In our motivating clinical trial, one of our baseline covariates is actually a pre-treatment measurement of the outcome $T.$ Because of this, we could choose to define the true outcomes several ways with the possible benefit of maximizing efficiency. For example, we could define the outcomes $T(0), T(1)$ as the original values of the endpoints and use $X$ for subgroup analysis. Alternatively, we could analyze the outcomes as the change from baseline measurement $X$ to later measurements, denoted as $T^D(0)$ and  $T^D(1)$. We would like to know if it is advantageous to define a trial outcome in terms of change from baseline compared to using the baseline value as a covariate. To be explicit about the quantities we are estimating, we can outline the relevant joint and conditional distributions based on the choice of endpoint definition and involvement of $X$. All four methods we consider are based on the joint 4x4 distribution from equation 1 extended to include $X$. In the special setting where our covariate is a pre-measurement of $T(0)$, we briefly consider other strong assumptions that could be made. For example, in the strongest case we could plug in $X$ in for $T(0)$ and gain identifiability. Alternatively, we could view $X$ and $T(0)$ as repeated measures and assume certain parameters are equal, such as identifiable means and variances or nonidentified correlations. We consider how these relate to conditional independence assumptions proposed by other authors in future work. 
 
In the setting we are considering, there is a pre-planned analysis of the final clinical trial data, and defining the endpoint is necessary to test for a treatment effect. Like we have suggested for assessing surrogacy, this step involves fitting marginal or conditional models with the original outcome or difference from baseline endpoint. For efficiency or to incorporate treatment effect subgroups, we may choose to condition on baseline covariates $X$. We enumerate the potential analysis models for the treatment effect based on the observed outcome $T$ and the corresponding surrogate validation models for the potential outcomes $T(0)$ and $T(1)$ conditional on $S(1) = s$. These are differentiated by the endpoint:
\begin{enumerate}
\item Original outcome: $T(0), T(1)$ \ \ \ \ Treatment effect model $T_i = \beta_0 + \beta_1 Z_i + \epsilon_i$
\vspace{-.1cm}

Surrogate validation metric $E(T(1) - T(0)|s) = \gamma_{0,M} + \gamma_{1,M}s$

\item Also condition on $X$:  $T(0), T(1)|X$ \ \ \ \ Treatment effect model 
$T_i = \beta_2 + \beta_3 Z_i + \beta_4 X_i + \epsilon_i$

Surrogate validation metric $E(T(1) - T(0)|X,s) = \gamma_{0,C} + \gamma_{1,C}s$

\item Difference from baseline: $T^D(0), T^D(1)$ \ \ \ \ Treatment effect model $T^D_i = \beta_5 + \beta_6 Z_i + \epsilon_i$

Surrogate validation metric $E(T^D(1) - T^D(0)|s) = \gamma_{0,M} + \gamma_{1,M}s$

\item Also condition on $X$: $T^D(0),T^D(1)|X$  \ \ \ \ \ Treatment effect model $T^D_i = \beta_7 + \beta_8 Z_i + \beta_9 X_i + \epsilon_i$

Surrogate validation metric $E(T^D(1) - T^D(0)|X,s) = \gamma_{0,C} + \gamma_{1,C}s$

\end{enumerate}
Parameters $\beta_1, \beta_3, \beta_6,$ and $\beta_8$ estimate the respective treatment effects. We expect the first and third methods to produce estimates of the same population, marginal treatment effect regardless of subtracting off the baseline measurement, though finite sample equality is unlikely to hold for even a randomized trial. These same considerations extend to the surrogate validation framework, where some of these methods will estimate the same marginal validation estimates. We will later consider a further reason to thoughtfully define the endpoint, which is to determine which scale it is reasonable to assume conditional independence. 

\section{Bayesian Methods}

\subsection{Imputation-Estimation Algorithms}

While we observe only $S(0),T(0)$ for $n_0$ subjects and $S(1),T(1)$ for $n_1$ subjects, our validation quantities involve correlations that can only be calculated from counterfactual outcomes. In order to simultaneously estimate the model parameters, address nonidentified terms, and to appropriately propagate the uncertainty of imputing missing outcomes, we use a Bayesian method for estimation. There are three types of variables that will be iteratively drawn in the MCMC algorithm: the correlation parameters in $R$, the model mean and variance parameters ($\mu$'s and $\sigma$'s), and the missing potential outcomes. The imputation strategy is a full process that iteratively imputes the missing potential outcomes and uses the posterior distribution to draw values of the parameters. 

We assume vague, normal priors for the identified mean parameters. Rather than use an Inverse-Wishart prior or another method that would sample the entire matrix at once that lacks needed flexibility, we implement a separation method on the covariance matrix $\Sigma$. This decomposes the matrix $\Sigma= QRQ$ into standard deviation and correlation matrices where $R$ is a correlation matrix with $1$'s on the diagonal to easily place less informative priors on the identified terms.$^{27}$ To ensure iterative draws satisfy the positive definite constraint on $\Sigma$, the method uses the griddy Gibbs algorithm to draw from the appropriate bounded posterior.$^{20}$ Specifically, we compute the posterior of each parameter over a set of realizable grid points and re-evaluate over a region of high posterior density with more precision (finer grid points) before randomly drawing the value for that iteration. We consider different priors for the correlation parameter and find that it is important to carefully choose on which parameters to place priors when implementing the conditional independence constraint. Since we are in a setting with nonidentified correlation parameters where the data will provide no direct information about the true values of these parameters, we are careful to not impose unreasonable prior distributions as we expect the posterior to mimic the prior. We consider both vague and more informative Uniform and Beta priors on the correlation terms, though the marginal distribution of each correlation under positive definite constraints can become less straightforward. We assume $S(1) \perp T(0 )| T(1)$ (or its equivalent based on the exact model fit and incorporation of $X$), by drawing suitable values of $\theta_T$ and $\theta_{11}$ using the grid search. Essentially, as demonstrated in equation 4, the term $\theta_{10}$ is no longer involved in the likelihood, and the product $\theta_T \times \theta_{11}$ takes its place.

\subsection{Observed Data Algorithm}

An alternative way to estimate the identifiable parameters is by using the observed data likelihood and devising an MCMC algorithm to obtain draws from the posterior distribution.  This approach avoids imputing the counterfactual values of $S$ and $T$. Since the data contain no information about $\theta_T$ or $\theta_{10}$, we expect the posterior to match the distribution of the chosen prior, provided the priors are independent. Thus, we propose to draw the nonidentified correlation parameters directly from the prior solely for the purposes of estimating $\gamma_0$ and $\gamma_1$.  Since the value of $\theta_{11}$ is estimable from the observed data, we use the posterior distribution for its draw, but it is the only correlation parameter drawn from a conditional distribution. When assuming conditional independence, we replace the term $\theta_{10}$ with $\theta_T \times \theta_{11}$ (after $\theta_{11}$ is drawn from the posterior and  $\theta_T$ is drawn directly from \\Uniform(-1, 1)) in the same way as explained above when calculating values of $\gamma_1$ and ensuring the matrix is positive definite. We then fit the regression models on the observed data only to estimate the other parameters.
   
Here we carry out basic Bayesian estimation of the identified parameters for comparability of uncertainty estimates to isolate the effect of using the imputation scheme and priors for nonidentified parameters. We find that we can bypass the full MCMC scheme intended to provide parameter estimates while addressing the nonidentified parameters, and instead we can draw these independently from the prior. This is related to work by Gustafson that demonstrates any difference between the prior and posterior distribution for non-identified parameters is due to prior dependence in the parameters.$^{28}$ Using his transparent reparameterization here, there is no indirect learning of the correlations outside of the positive definite and conditional independence assumption constraints, which are still enforced with this algorithm. In this setting, we expect results from this method to be generally equivalent to imputation while being less computationally expensive.  We also note that using an MCMC scheme to estimate the variance and mean parameters may not be necessary at all, and a maximization of the posterior or maximum likelihood method may be used instead.

\section{Simulation Studies}

We explore the impact of using a baseline covariate in terms of efficiency and making conditional independence assumptions. In particular, our simulations are meant to assess how we define the true endpoint based on different relationships with the baseline covariate. Using simulation studies, we generate data that mimics a randomized trial, meaning for half of the subjects assigned $Z=0$, we observe only $T(0), X$, and for the other half we observe $S(1), T(1), X$. The five sets of generative parameters for equation 2 are shown in Table 1. The five settings have different generating parameters that vary the treatment effect and quality of the surrogate endpoint to demonstrate the method's performance over a variety of scenarios. Based on the model in equation 2 for observed data of sample size $n = 100$, we generate data that fulfills conditional independence in all scenarios using the original endpoint and conditional on $X$ setting. [\autoref{tab_gener_values}]

The surrogate is valid marginally only for settings $A, B$. The covariate $X$ is normally distributed in settings $A$-$D$ and binary in $E$; settings $D,E$ represent the existence of a subgroup effect. After generating the data, we fit the models described below and vary which conditional independence assumption is made, if any. To perform surrogacy validation, we fit four models of tri-variate normal distributions derived from the distribution of $S(1), T(0), T(1), X$ (see Web Appendix B for details). The four analysis designs are based on the distribution of the three outcomes and baseline covariates, and the different parameterizations can be easily equated algebraically. For any $X$ (we simulate $X \sim N(\delta_4, \sigma_X^2)$ or $X \sim Bernoulli(0.5)$),

\vspace{0.25cm}

\begin{enumerate*}[series=MyList, before=\hspace{-0.6ex}]

\item $ \left(
\begin{array}{c}
S(1) \\ T(0)  \\ T(1) 
\end{array}
\right) $  

\item $ \left(
\begin{array}{c|c}
S(1)  & \\ T(0)  &X \\ T(1)  &
\end{array}
\right) $

\item
$ \left(
\begin{array}{c}
S(1)  \\ T^D(0) \\ T^D(1) 
\end{array}
\right) $ 

\item  $ \left(
\begin{array}{c|c}
S(1)  &  \\ T^D(0)  &X \\ T^D(1) &
\end{array}
\right) $  
\end{enumerate*}

\vspace{.35cm}
 We consider that conditional on baseline covariates based on context, we may decide to introduce strong modeling assumptions. For example, when $X$ is a pre-treatment measurement of $T$ (like a measurement-error prone value of $T(0)$), priors may be informed by estimates of the observed correlation of $X$ and $T(1)$, or we may be able to estimate some nonidentified correlations. Currently we use a Beta prior (truncated between -0.4 to 1 with a positive mean equal to 0.23) on the correlation between either $T(0), T(1)$ or $T^D(0), T^D(1)$ and a Uniform(-1, 1) prior on the correlation between $S(1), T(1)$  or $S(1), T^D(1)$ when conditional independence is not assumed. To perform sensitivity analyses, we will both vary the prior distributions on the nonidentified parameters to integrate over the range of plausible values and fix the correlations to see at what boundaries the conclusions change. 

These models allow us to contrast the estimation of marginal quantities $\gamma_{0, M}$ and $\gamma_{1, M}$ and also those for subgroups of participants based on baseline covariates $X$ from models 2 and 4. For our purposes, we first compare the marginal estimates for $\gamma_1$ and $\gamma_0$ which are directly calculated in settings 1 and 3 and are derived by marginalization for models 2 and 4. The definition of the endpoints (either the difference from baseline $D$ and corresponding $\gamma_{.D}$ or the original value $O$ and $\gamma_{.O}$) is important as $\gamma_{1, D} \neq \gamma_{1, O}$ when there is an effect of $X$. Further, the validity of conditional independence varies: when $T(0)\perp S(1)|T(1), X$, it is also true that $T(0)^D \perp S(1) | T(1)^D, X.$ However, this corresponding satisfaction of conditional independence does not necessarily hold between $T(0) \perp S(1) | T(1)$ and $T(0)^D \perp S(1) | T(1)^D$ when we do not condition on $X$. To quantify this discrepancy when this condition is not met, we calculate the deviation from meeting the conditional independence requirement, i.e. how incorrect it is to make this conditional independence assumption (see Web Table 1 in the supporting information).

\vspace{-.22in}
\subsection{Simulation Results}

We present the posterior mean point estimates averaged over 100 datasets using the observed data algorithm. To assess variability, we also report the standard deviation of the Bayesian point estimates and the average standard error of the parameter value (the posterior standard deviation) within each replication. We run each MCMC for 3,000 iterations and ensure convergence is reached using traceplots. Since our goal is to effectively validate surrogate endpoints, we focus on inference for the quantities $\gamma_0$ and $\gamma_1$. The results show the contrast in estimation accuracy and efficiency when the constraints are enforced during estimation compared to when no assumptions are imposed during the MCMC procedure. 

\subsection{Marginal Estimates}

\begin{center}
\begin{figure}[H]
\caption{Simulation results for the two definitions of the endpoint (here categorized by superscript to denote $O$ = Original Endpoint (Settings 1 and 2) compared to $D$ = Difference from baseline (Settings 3 and 4)) and different generating parameter values $(A-E)$. The values shown below are the bias and variability (both the average within-sample standard error across datasets and the standard deviation of the point estimates) of the validation quantities that are adjusted by the variability of the hypothetical surrogate values if all counterfactuals were to be observed (from the full data). }
\vspace{-.2cm}
\begin{center}
\includegraphics[width = 6.3in]{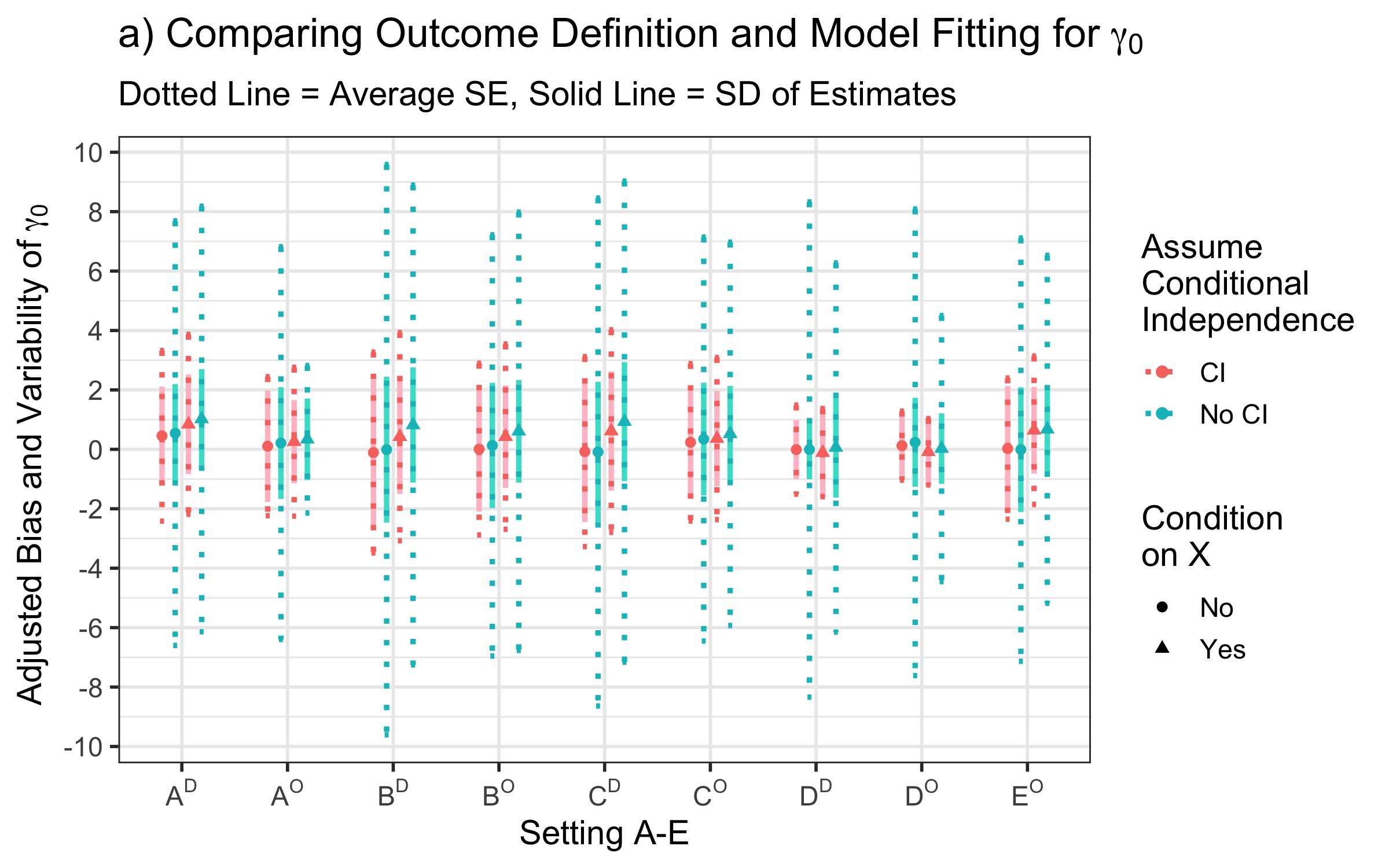}
\includegraphics[width = 6.3in]{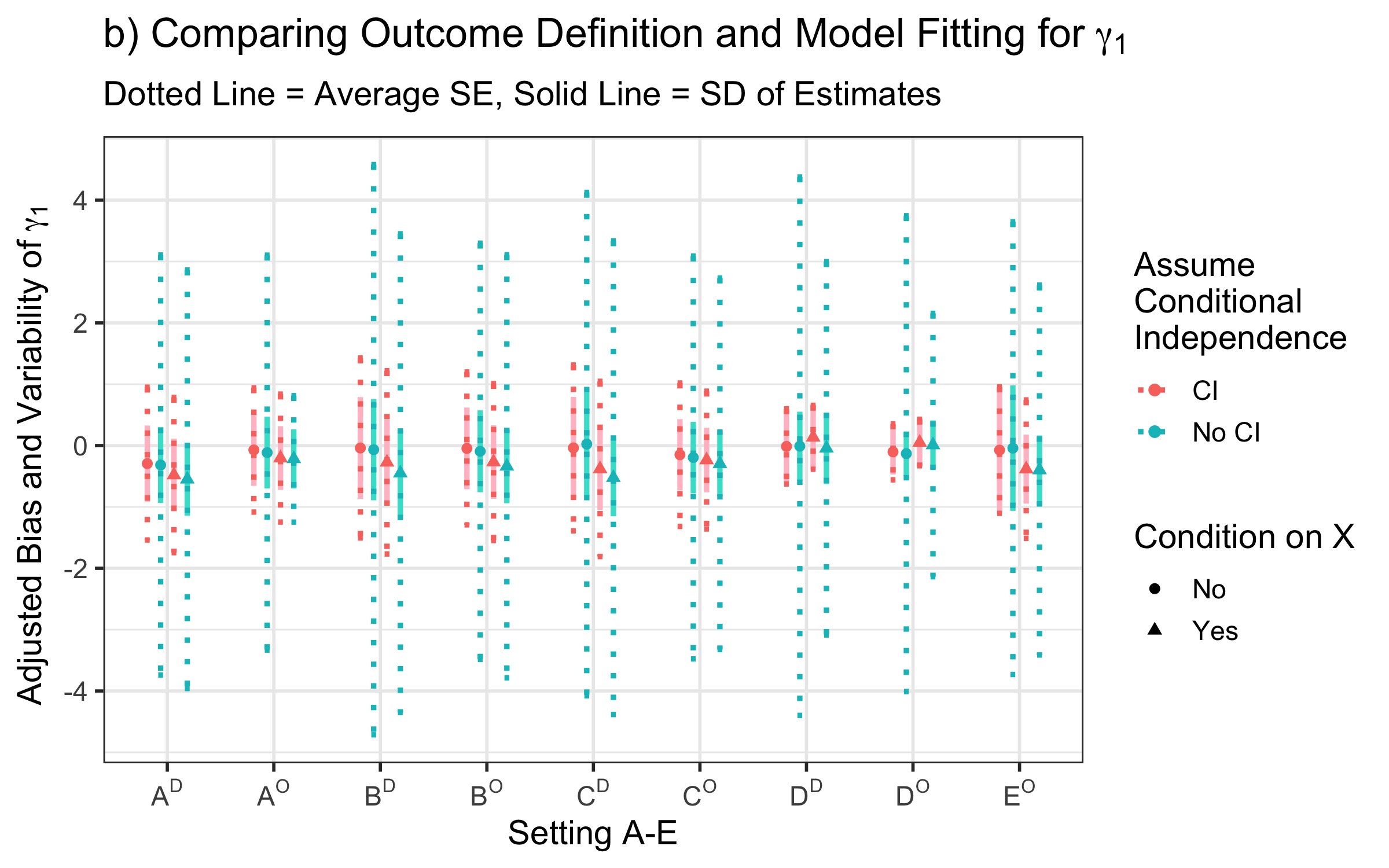}
\end{center}
\label{plotsf1}

\end{figure}

\end{center}

Below [\autoref{plotsf1}] (and in Web Table 1) are results  with sample size $n = 100$ using a truncated Beta($5,6$) prior. Bias for rows 1 and 2 as well as 3 and 4 is calculated based on the same true, marginal values. Overall, the identified mean and variance parameters are estimated with little bias. However, the nonidentified correlations are sensitive to the prior distribution. Compared to Web Table 1, for comparability across settings Figure 1 has been adjusted by the empirical variability in the validation estimates from complete counterfactual data: we calculated the maximum likelihood estimates under a scenario where we would observe all $n$ counterfactual outcomes and fit a model for  $T(1) - T(0)$ conditional on $S(1)$. We adjusted the bias and standard errors relative to the standard deviation of the estimates across 100 simulations to provide quantities that can interpreted in proportion to the amount of variability in the data and the estimates.

These results show there is often reduced bias in the estimates when making conditional independence assumptions depending on the exact setting and prior specification.  Notably, the credible intervals for the nonidentified parameters are very conservative for all settings as expected due to the non-identified correlation parameter and its associated relatively weak prior, and the corresponding coverage probabilities are near one. Further, the within simulation average, over-estimation of the standard error decreases when making conditional independence assumptions, though the SD of the estimates is not necessarily smaller when making these assumptions (through less under-identification in the conditional independence model, this reduction in SE improves the agreement between the average SE and SD). We see that conditioning on $X$ can also decrease the standard error of the estimates, particularly in the scenario where we do not make conditional independence assumptions. 

\subsection{Sensitivity to Distributional Assumptions}

We show results in supporting information Web Table 2 for fitting conditional models when the outcomes are heavy tailed (t-distributed) or skewed (gamma-distributed). When data follows these non-Gaussian distributions, the estimation results do vary from previous simulations. While there is some increased bias and variability of the estimates as compared to those with a normal distribution, generally there is some robustness in the estimation such that the conclusions regarding surrogate validity seem to be similar in the settings we considered.

\subsection{Subgroup Analysis}

Now we focus on the results of simulation setting $E$ where subgroups exist for the models fit using methods 2 and 4 (conditional on $X$). Since $X$ is binary in this setting, we report the conditional values of $\gamma_0$ when $X=0$ and $X=1$. [\autoref{sgroups}] Whereas Figure 1 shows marginalized estimates of the quantities, Table 2 shows the estimates of $\gamma_0$ and $\gamma_1$ conditional on the values of $X$, which allows us to determine if the surrogate is valid for patients with certain baseline characteristics. Looking at the validity of $S$ as a surrogate for subgroups in setting $E$, we see that the surrogate is valid for $X=0$ but not for $X = 1$. If $X$ were gender, for example, this would indicate that the surrogate is valid only for males and not females. However, this is only detected for the model where we make a conditional independence assumption. When we do not, the credible interval of $\gamma_1$ overlaps with 0 in almost all simulation replications, signaling that we are not able to declare the surrogate as valid.

\vspace{-.5cm}
\section{Duchenne Muscular Dystrophy Data Example}

We generate ``observed" data for three normally-distributed outcomes aimed to mimic an ongoing clinical trial and the natural progression and deterioration of ambulatory function for Duchenne muscular dystrophy patients.$^{29}$ Since the trial has not been unblinded, the parameter values were chosen to match the preliminary data when available, and for other parameters the values were chosen in a subjective way to be what the authors considered as reasonable. Based on the literature and company analysis standards, to assess surrogacy, we condition on age at baseline, $A$, and the baseline NSAA score measurement, $X$.


\subsection{Data Example Results}
\begin{figure}[H]
\begin{center}

\caption{The CEP plot shows the conditional functions of $\gamma_0$ and $\gamma_1$ across possible values of $S(1)$. The empirical distribution of $S(1)$ is shown in the blue density curve.}.
\includegraphics[width = 4in]{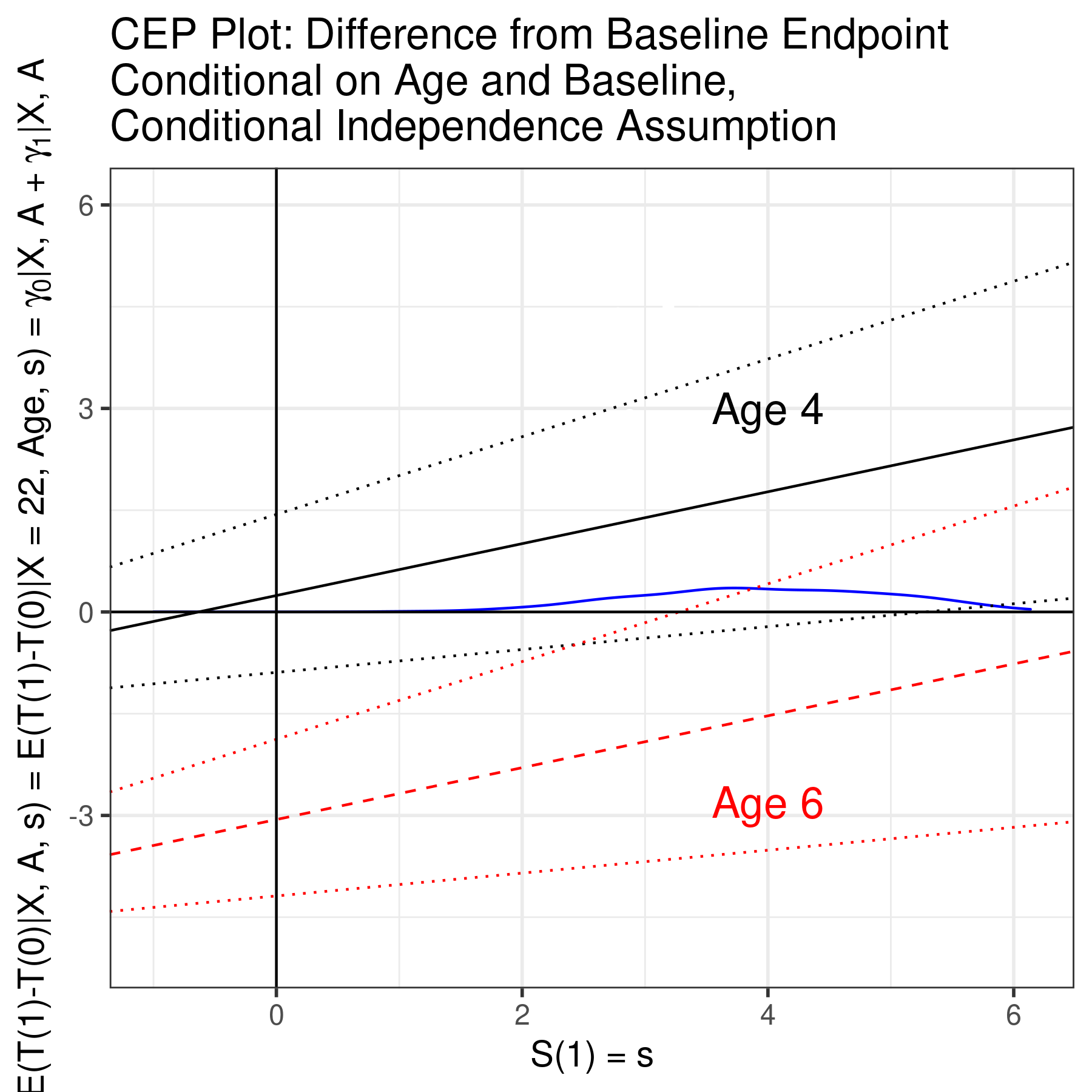}
\label{cepf2}

\end{center}
\end{figure}

First we show the estimated treatment effects that would be calculated in a clinical trial measuring the efficacy of the treatment based on the true, observed endpoints $T$. [\autoref{tab_duchenne_ex}] It is clear that the standard error estimates for $\gamma_0$ and $\gamma_1$ are markedly smaller, and credible intervals more narrow, when we make the conditional independence assumptions. We can see that marginally, the surrogate of micro-dystrophin is not a valid surrogate for improvement of the NSAA score across the entire study sample. However, there are strong effects of both age and baseline measurement on the outcome. We can identify a region of the covariate space based on age and baseline measurement where the surrogate is valid for a subgroup of patients. We create CEP curves for these data to show what the surface looks like when we stratify based on covariates. [\autoref{cepf2}]

The CEP plot is conditional on baseline NSAA and age, and we demonstrate that the surrogate is valid for those at four years of age. Since age is modeled with linear and quadratic effects, the surrogate will be invalid for those six and older as the estimated CEP curve  moves farther from 0. This demonstrates that due to the combination of natural growth and degeneration due to disease over time, the surrogate would only be valid within a certain younger patient population.

We also explore the consequences of different prior distributions for the non-identified parameter $\theta_T$ and compare this to fixing $\theta_T$ at some value, shown in Web Figure 1. We see that the results for $\gamma_0$ and $\gamma_1$ are somewhat sensitive to the choice, but the conclusion that the surrogate is valid holds for values of $\theta_T$ which we believe to be reasonable. The results are the same for both the imputation and observed data algorithms.

\vspace{-.2in}

\section{Discussion}

In this work, we have focused on incorporating baseline covariates into the validation process for surrogate endpoints. Considering the harmful implications of incorrectly validating a surrogate endpoint, it may be worthwhile to consider the CEP as a function of $X$ and identify potential subgroups of patients for which a surrogate is appropriate. It is important to assess the plausibility of conditional independence assumptions even in this context of adjusting for baseline covariates which we believe makes the assumptions more likely. Introducing these conditional independence constraints aids in improved estimation properties, but they should not be implemented without proper, context-dependent reasoning. Further, there are many ways to implement the constraints, although these simulations suggest certain strategies (using observed data only) may reduce the burden of imputing potential outcomes but still require reasonably well-specified prior distributions. 

There are many ways in which these methods and simulations will be extended, particularly to verify its robustness. Work is ongoing to incorporate baseline covariates and conditional independence assumptions into this framework while fully utilizing the longitudinal data.
\vspace{-.2in}

\section*{Data Availability Statement}

Data sharing is not applicable to this article as no new data were created or analyzed in this study.
\vspace{-.5cm}
\section*{Acknowledgements}

This work was supported by the National Science Foundation (DGE 1256260) to ER; the NIH (CA129102) to JMGT. Thanks to Walter Dempsey for useful discussions.

\newpage

\section* {References} 
\begin{enumerate}

\item Prentice R. Surrogate endpoints in clinical trials: definition and operational criteria. \textit{Statistics in Medicine.} 1989; {8(4),}  431-40.

\item VanderWeele TJ.  Surrogate measures and consistent surrogates. \textit{Biometrics.} 2013; { 69(3),} 561-565.

\item Rubin DB. Estimating causal effects of treatments in randomized and nonrandomized studies.  \textit{Journal of Educational Psychology.} 1974; {66(5),}    688-701.

\item  Little R, Rubin D. Causal effects in clinical and epidemiological studies via potential outcomes: concepts and analytical approaches. \textit{ Annual Review of Public Health.} 2000; {21(1),} 121-145.

\item Frangakis C, Rubin D. Principal stratification in causal inference. \textit{ Biometrics.} 2002; {  58(1),}  21-29.

\item Gilbert P,   Hudgens M. Evaluating candidate principal surrogate endpoints. \textit{ Biometrics.} 2008; {  64(4),} 1146-1154.

\item Mendell JR, Sahenk Z, Lehman K, Nease C, Lowes LP, Miller NF, et al. Assessment of Systemic Delivery of rAAVrh74. MHCK7. micro-dystrophin in Children With Duchenne Muscular Dystrophy: A Nonrandomized Controlled Trial. \textit{ JAMA Neurology.} 2020; 77(9), 1122-1131.

\item Halloran ME, Pr\'eziosi MP,  Chu H. Estimating vaccine efficacy from secondary attack rates.  \textit{Journal of the American Statistical Association.} 2003; { 98(461),} 38-46.

\item  Pr\'eziosi MP,  Halloran ME. Effects of pertussis vaccination on disease: vaccine efficacy in reducing clinical severity. \textit{Clinical Infectious Diseases.} 2003; {37(6),}  772-779.

\item  Hudgens MG,  Halloran ME.  Causal vaccine effects on binary postinfection outcomes. \textit{  Journal of the American Statistical Association.} 2006; { 101(473),} 51-64.

\item Follmann D. Augmented designs to assess immune response in vaccine trials. \textit{ Biometrics.} 2006; {  62(4),} 1161-1169.

\item Gabriel EE,  Gilbert PB. Evaluating principal surrogate endpoints with time-to-event data accounting for time-varying treatment efficacy.  \textit{ Biostatistics.} 2014; {  15(2),} 251-265.

\item Gabriel EE,  Follmann D. Augmented trial designs for evaluation of principal surrogates. \textit{  Biostatistics.} 2016; { 17(3),} 453-467.

\item Gilbert PB, Qin L, Self SG.  Evaluating a surrogate endpoint at three levels, with application to vaccine development. \textit{  Statistics in Medicine.} 2008; {  62(4),} 4758-4778.

\item  Wolfson J,  Gilbert P.  Statistical identifiability and the surrogate endpoint problem, with application to vaccine trials. \textit{Biometrics.} 2010; { 66(4),}  1153-1161.

\item  Huang Y, Gilbert PB,  Wolfson J. Design and estimation for evaluating principal surrogate markers in vaccine trials. \textit{ Biometrics.} 2013; {  69(2),}  301-309.

\item Zhuang Y, Huang Y,  Gilbert PB.  Simultaneous Inference of Treatment Effect Modification by Intermediate Response Endpoint Principal Strata with Application to Vaccine Trials. \textit{The International Journal of Biostatistics.} 2019; {16(1).}

\item Gilbert PB,  Huang Y. Predicting overall vaccine efficacy in a new setting by re-calibrating baseline covariate and intermediate response endpoint effect modifiers of type-specific vaccine efficacy. \textit{ Epidemiologic Methods.} 2016; {  5(1),}  93-112.

\item  Zigler CM,  Belin TR. A Bayesian approach to improved estimation of causal effect predictiveness for a principal surrogate endpoint. \textit{Biometrics.} 2012; {68(3),} 922-932.

\item  Conlon AS, Taylor JM, Elliott MR. Surrogacy assessment using principal stratification when surrogate and outcome measures are multivariate normal.  \textit{ Biostatistics.} 2014;
{15(2),}  266-283.

\item Alonso A, Van der Elst W,  Meyvisch P. Assessing a surrogate predictive value: a causal inference approach.  \textit{Statistics in medicine}. 2017; {36(7)}, 1083-1098.

\item  Conlon A, Taylor J, Li Y, Diaz-Ordaz K,  Elliott M. Links between causal effects and causal association for surrogacy evaluation in a gaussian setting. \textit{ Statistics in Medicine.} 2017;
{ 36(27),} 4243-4265.

\item  Taylor JM, Conlon AS,  Elliott MR. Surrogacy assessment using principal stratification with multivariate normal and Gaussian copula models. \textit{Clinical Trials.} 2015; {12(4),}  317-322.

\item Ding P, Feller A,  Miratrix L. Randomization inference for treatment effect variation. \textit{ arXiv preprint.} 2014; arXiv:1412.5000.

\item  Parast L, Cai T, Tian L.  Nonparametric estimation of the proportion of treatment effect explained by a surrogate marker using censored data. \textit{Technical Report.} 2016.

\item  Parast L, McDermott MM, Tian L. Robust estimation of the proportion of treatment effect explained by surrogate marker information.
\textit{Statistics in Medicine.} 2016; {35(10)} 1637-1653.

\item  Barnard J, McCulloch R,  Meng XL.  Modeling covariance matrices in terms of standard deviations and correlations, with application to shrinkage.  \textit{Statistica Sinica}. 2000; 1281-1311.

\item Gustafson P. What are the limits of posterior distributions arising from nonidentified models, and why should we care?. \textit{Journal of the American Statistical Association}. 2009; {  104(488)}, 1682-1695.

\item  Muntoni F, Domingos J, Manzur AY, Mayhew A, Guglieri M, UK NorthStar Network, et al. Categorising trajectories and individual item changes of the North Star Ambulatory Assessment in patients with Duchenne muscular dystrophy. \textit{PloS One.} 2019;{   14(9),} e0221097.

\end{enumerate}

\newpage
\subsection{Tables and Figures}

\begin{table}[H]
\begin{center}
\resizebox{\textwidth}{!}{
\begin{tabular}{ l ccccc }
 & A & B & C & D & E\\ \hline \hline
 & Valid $S$ & Valid $S$: No X Effect & Invalid $S$  & Valid $S$ for SG  & Valid $S$ for SG \\ \hline
\rowcolor{lightgray} $\sigma_X$ & 0.5 & 0.5 & 0.5 & 0.5 & NA \\
$\delta_4$ & 1 & 1 & 1 & 1 &  NA \\
\rowcolor{lightgray} $\omega_1$ & 2 & 2 & 2 & 2 & 2 \\
$\omega_2$ & 0 & 0 & 0 & 0 & 0 \\
\rowcolor{lightgray}
$\omega_3$ & 3 & 3 & 3 & 3 & 3 \\
$\omega_4$ & 1 & 0 & 1 & 3 &-0.75 \\
\rowcolor{lightgray} $\omega_5$ & 4.1 & 4.1 & 4.1 & 4.1 & 4.1 \\
$\omega_6$ & 1 & 0 & 1 & 1 & 2 \\
\rowcolor{lightgray} $\epsilon_{S1} = \epsilon_{T0} = \epsilon_{T1}$ &  1 &  1 &  1 &  1 &  1\\
$\theta_{10}$ & 0.15 & 0.15 & 0.15 & 0.08 & 0.15 \\
\rowcolor{lightgray} $\theta_{11}$ & 0.7 & 0.7 & 0.7 & 0.3 & 0.7 \\
$\theta_{T}$ & 0.21 & 0.21 & 0.21 & 0.26  & 0.21 \\
\rowcolor{lightgray} $\gamma_{0, O}$ & 0 & 0 & -1.00 & -1.35 & 1.31 \\
$\gamma_{1, O}$ & 0.55 & 0.55 & 0.55 & 0.22  & 0.58 \\
\rowcolor{lightgray} $\gamma_{0, D}$ & -0.06 & 0  & -1.02 & -1.33 & NA\\ 
$\gamma_{1, D}$ & 0.58 & 0.55 & 0.56 & 0.22 & NA \\
 \hline \hline
\end{tabular}
}
 \caption{Generative parameter values for the five scenarios to compare definition of the endpoint and using baseline covariates. SG stands for subgroup based on $X$. The meaning of these parameters is shown in equation 2. Note `difference from baseline' (subscript $D$) is not defined when $X$ is binary and $T$ is continuous ($E$). }
    \label{tab_gener_values}
\end{center}
\end{table}

\begin{center}
\hspace{-.2cm} \begin{table} [H]
\begin{singlespace}

\resizebox{.78\textheight}{!}{
\begin{tabular}{ ll c c c c c  c c c c c c c c c c }
Setting & Fit Conditional & $\gamma_{0, C}$  &  &   && & $\gamma_{0, C}$  & &  &   && $\gamma_{1, C}$ &      \\
& Independence & True & Est & SE & SD & Covers & True & Est & SE & SD & Covers & True & Est & SE & SD & Covers  \\
& Assumption & $X = 0$&&&& 0 &$X = 1$ &&& & 0 & && && 0 \\
\hline
2E  &  $T(0) \perp S(1) |  T(1), X$ & 0 & 0.064 & 0.472 & 0.308 & 1.00 &2.75 & 2.778 & 0.470 & 0.312 & 0.00 & 0.55& 0.512 & 0.179 & 0.089 & 0.00 \\
2E  & None & 0 & 0.078 & 1.024 & 0.309 & 1.00 & 2.75 & 2.784 & 1.013 & 0.312 &  0.04 & 0.55 & 0.509 & 0.481 & 0.089 & 0.98 \\
\hline
\hline 
\end{tabular}
}
\end{singlespace}
 \caption{Simulation results demonstrating effect of estimating subgroups. Estimates of $\gamma_0$ and $\gamma_1$ are conditional on the values of $X$, so values of $\gamma_0$ estimates are conditional on $X = 0,1$. The columns denoting `covers 0' indicate what proportions of simulations have credible intervals that do contain 0. This helps determine for how many simulations the surrogate would be considered valid. For $\gamma_0$, the credible interval covering 0 denotes a valid surrogate, while the interval of $\gamma_1$ should not.}
    \label{sgroups}
    \end{table}
    \end{center}

\begin{center}
\begin{table} [H]
\resizebox{\textwidth}{!}{
\begin{tabular}{ c  c  c  c  c  c  c }
Endpoint  & Treatment & Conditional Independence  & $\gamma_0$ & $\gamma_0$  & $\gamma_1$ &$\gamma_1$  \\
Type  & Effect & Assumption & Estimate & SE & Estimate & SE \\
 \hline
$T^D$  & 0.227 &  $T^D(0) \perp S(1)| T^D(1),X,A$ &  -3.040 & 0.393 & 0.393 &  0.101  \\
 && None & -2.980 & 2.235  &  0.386 & 0.279 \\ \hline

\hline
$T$  & 0.271 &  $T(0) \perp S(1)| T(1),X,A$ &   -3.046  & 0.826  &  0.394 &  0.102 \\
&&  None & -2.986 & 2.232  &  0.386 & 0.279  \\ \hline

\end{tabular}
}
 \caption{Simulated Muscular dystrophy estimates of treatment effect and marginal surrogacy quantities.}
    \label{tab_duchenne_ex}
\end{table}
\end{center}

\vspace{-1.5cm}

\vspace{-1.7cm}

\end{document}